\documentclass[useAMS,usenatbib]{mn2e}
\usepackage[totalwidth=515pt,totalheight=660pt,left=1.4cm,right=1.4cm]{geometry}
\usepackage{graphicx,amssymb,color}
\usepackage[normalem]{ulem}
\input psfig

\title[Post-MS rotational fission of asteroids]
{Post-main-sequence debris from rotation-induced YORP break-up of small bodies}
\author[Veras, Jacobson \& G\"{a}nsicke]{
Dimitri Veras$^{1}$\thanks{E-mail: d.veras@warwick.ac.uk},
Seth A. Jacobson$^{2,3}$,
Boris T. G\"{a}nsicke$^{1}$
\\
$^{1}$Department of Physics, University of Warwick, Coventry CV4 7AL, UK
\\
$^{2}$Laboratoire Lagrange, Observatoire de la C\^{o}te d'Azur, Boulevard de l'Observatoire, F-06304 Nice Cedex 4, France
\\
$^{3}$Bayerisches Geoinstitut, Universt\"{a}t Bayreuth, D-95440 Bayreuth, Germany
}

\begin{document}

\date{Accepted 2014 September 14. Received 2014 September 08; in original form 2014 July 24}

\pagerange{\pageref{firstpage}--\pageref{lastpage}} \pubyear{XXXX} 

\maketitle

\label{firstpage}

\begin{abstract}
Although discs of dust and gas have been observed orbiting white dwarfs, the origin of this circumstellar matter is uncertain.  We hypothesize that the in-situ breakup of small bodies such as asteroids spun to fission during the giant branch phases of stellar evolution provides an important contribution to this debris.  The YORP effect, which arises from radiation pressure, accelerates the spin rate of asymmetric asteroids, which can eventually shear themselves apart.  This pressure is maintained and enhanced around dying stars because the outward push of an asteroid due to stellar mass loss is insignificant compared to the resulting stellar luminosity increase.  Consequently, giant star radiation will destroy nearly all bodies with radii in the range 100 m - 10 km that survive their parent star's main sequence lifetime within a distance of about 7 au; smaller bodies are spun apart to their strongest, competent components.  This estimate is conservative, and would increase for highly asymmetric shapes or incorporation of the inward drag due to giant star stellar wind.  The resulting debris field, which could extend to thousands of au, may be perturbed by remnant planetary systems to reproduce the observed dusty and gaseous discs which accompany polluted white dwarfs.
\end{abstract}

\begin{keywords}
minor planets, asteroids: general -- stars: white dwarfs -- methods: numerical -- 
celestial mechanics -- planet and satellites: dynamical evolution and stability
-- protoplanetary discs
\end{keywords}

\section{Introduction}

Observations unambiguously demonstrate that single white dwarfs (WDs) harbour dusty and gaseous 
circumstellar discs.  The mounting discoveries, particularly within the last decade, result in a current
tally of about 30 dusty discs \citep{zucbec1987,becetal2005,kiletal2005,reaetal2005,faretal2009} 
and 7 gaseous discs \citep{ganetal2006,ganetal2007,ganetal2008,gansicke2011,faretal2012,meletal2012}; 
some systems include both dusty and gaseous discs \citep{brietal2009,meletal2010,brietal2012}.
Every one of these discs is accompanied by detections of metals in WD atmospheres, indicating accretion from the disc. 

Although we know that the disc matter arises from remnant planetary systems, we do not know from what part of those systems.  Whether the progenitors, which remain undetected, are planets, asteroids or comets, they are assumed to be tidally or sublimationally disrupted into discs \citep{graetal1990,jura2003,beasok2013,stoetal2014,veretal2014a}. Dust modeling provides partial constrains on the particle sizes in these discs by yielding a lower limit on the order of a micron, but no upper limit \citep{reaetal2009}. Because the discs are extremely compact in
radial extent ($\sim 1 R_{\odot}$), planets have difficulty reaching such close separations.
A single planet's orbit will be pushed outward by mass loss during the giant branch
phases of stellar evolution, and can reverse direction
only by giant-star/planet tides \citep{musvil2012,adablo2013,norspi2013,viletal2014}, severe anisotropy in 
the mass-loss distribution \citep{veretal2013a}, or through stellar flybys \citep{veretal2014b}.  
All these scenarios are unlikely, as tidal prescriptions would have to be finely-tuned to
prevent engulfment into the star, mass loss can be considered isotropic within a few hundred au,
and the probability of a flyby causing a near-collisional orbital perturbation is too low.

Systems with multiple planets present different difficulties. Because mass loss changes the stability
boundary in multi-planet systems \citep{debsig2002,voyetal2013}, giant branch evolution
can trigger gravitational scattering instabilities amongst the planets.  Detailed simulations of 
two-planet \citep{veretal2013a} and three-planet \citep{musetal2014} systems across all phases
of evolution, including the main sequence and WD phases, indicate that the frequency of planets
scattered to near-collisional orbits with the WD is too small to explain the observations.

Smaller bodies transported inward from an external reservoir represent more likely progenitors.  Both compositionally and dynamically, comets 
cannot be the primary precursor of the discs \citep{zucetal2007,veretal2014c}, although comets can contribute to 
some discs, particularly around the youngest WDs \citep{stoetal2014}.  

Asteroids\footnote{Here and throughout the text, asteroid is used synonymously for small bodies and does not necessarily imply a rocky composition.} originating from an external belt are the more
likely choice.  By interacting with one planet, a fraction of a belt of debris can scatter close to or directly into the WD.  
\cite{bonetal2011} showed how a planet plus a Kuiper belt analogue can scatter some of
the belt into inner regions of a WD system.  \cite{debetal2012} demonstrated that
a planet can trap exo-asteroid belt members in an interior mean motion resonance 
before eventually scattering them to the Roche radius of the WD.  \cite{frehan2014}
extended these studies and found that when the planet is of a lower mass and eccentric,
then more asteroids encounter the WD radius. 

In summary, although explorations of the potential sources of disc progenitors in WD systems are at an early stage, asteroids 
are likely to be the dominant source.  This paper proposes that many asteroids will not survive the giant branch stages of 
stellar evolution intact due to radiation-induced rotational fission.  Consequently, the reservoir of material available to 
form discs around WDs is composed of smaller fragments than what was previously assumed.  We describe this type of disruption in Section 2 and 
conclude in Section 3.

\section{Spin evolution}

In this section, we first describe the primary driver of spin (Subsection \ref{sub1}) and
the critical spin at which disruption occurs (Subsection \ref{sub2}) before characterising
the spin evolution along the main sequence (Subsection \ref{sub3}), giant branches (Subsection 
\ref{sub4}) and WD (Subsection \ref{sub5}) phases of stellar evolution.  We conclude with numerical simulations 
(Subsection \ref{sub6}).

\subsection{The YORP effect} \label{sub1}

The YORP (Yarkovsky-O'Keefe-Radviesvki-Paddock) effect is the rotational acceleration of a body due to 
the anisotropic absorption and radiation of light. The effect was first proposed in the context of absorbing 
stellar UV and visible light and the emittance of thermal radiation due to the local heating of the 
body \citep{rubincam2000}, although the role of conducted heat is now considered as possibly as important 
\citep{golkru2012}. 

Observationally-measured rotational acceleration of a number of asteroids has matched predictions from 
the YORP theory.  Examples include 54509 YORP (2000 PH5) \citep{lowetal2007,tayetal2007} and 
1862 Apollo \citep{kaaetal2007}.  The YORP effect also torques the spin pole of small bodies, aligning 
prograde orbiters and anti-aligning retrograde orbiters with the angular momentum vector of their heliocentric orbits 
\citep{vokcap2002} and leaving entire asteroid families aligned \citep{voketal2003}, thereby
effecting the drift of smaller members due to the Yarkovsky effect \citep{botetal2002}. 
As the spin axis is re-oriented, the body is spun down or up, helping to explain both the excess 
of slow rotators in the asteroid population and the pile-up of rapid rotators at the spin barrier 
\citep{praetal2008,rosetal2009}.  Acceleration is also responsible for driving small bodies 
to rotational fission---the process whereby objects with little or no tensile strength can fall apart and
their elements enter a mutual orbit due to centrifugal accelerations matching or exceeding 
gravitational accelerations \citep{scheeres2007a,waletal2008,jacsch2011}. YORP-induced rotational 
fission leads to the creation of asteroid pairs \citep{waletal2008,praetal2010} and binary asteroids 
\citep{jacsch2011} which match the characteristics of those observed in the Solar System. 
This fission also significantly effects the size-frequency distribution of the asteroids 
in the Main Belt \citep{jacetal2014}.

Consider a solid body in orbit around a star.
The evolution of the spin ($s$) of this object may be expressed as \citep{scheeres2007b} 

\begin{equation}
\frac{ds}{dt} = 
        \frac{Y}{2 \pi \rho R^2}
\left(
        \frac{f}{a^2 \sqrt{1 - e^2}}
\right)
\label{sdef}
\end{equation}

\noindent{}where $R$ is the mean radius of the body, $\rho$ is its density, $a$ and $e$ are the semimajor axis
and eccentricity of its orbit, $f$ is a force and $Y \in [0,1)$ is a constant determined by the physical properties
of the body. The value of $Y$ may be thought of as the extent of asymmetry of the asteroid.  For a perfectly symmetric body, 
$Y = 0 = ds/dt$, which means that the asteroid's initial spin angular momentum is conserved and the asteroid maintains
its original spin.  Some Solar system-based examples 
include the asteroids 1862 Apollo, with $Y = 0.022$ \citep{kaaetal2007} and 54509 YORP, with $Y = 0.005$ \citep{tayetal2007}.  
Observational limitations currently prevent us from measuring $Y$ for small bodies residing outside of the Solar System.

The force $f$ is given by equations (23-26) of \cite{scheeres2007b} and is a function of
the incident stellar light pressure, the body's albedo, the fraction of radiation pressure due to specular reflection, 
and the Lambertian scattering coefficient.  For the Solar System, this force ($f = f_{\odot}$) is linearly proportional to the 
Solar radiation constant, and is approximately equal to $10^{17}$kg m/s$^2$.  Therefore, we assume here
for general planetary systems $f = k f_{\odot}$ where $k = L_{\star}/L_{\odot}$, such that $L$ denotes luminosity.  

\subsection{The critical disruption spin}  \label{sub2}

Observations of Solar System asteroids exhibit a sharp and unmistakable cutoff on a rotation period/radius diagram
(originally Fig. 1 of \citealt*{harris1994} and recently Fig. 1 of \citealt*{jacetal2014}), suggesting that asteroids 
with radii greater than about 250 m break up when their
rotational period becomes shorter than 2.33 hours.  The critical
spin at which disruption occurs ($\equiv s_{\rm crit}$) can take one of several analytical forms 
\citep[e.g., equations 34-38 of][]{davidsson1999}.  Given our lack of knowledge
of extrasolar asteroids, we assume they are strengthless rubble piles 
similar to what is observed in the Solar System \citep{ricetal2002}, and hence

\begin{equation}
s_{\rm crit} \equiv \frac{2\pi}{\sqrt{3\pi/G\rho}} = 
7.48 \times 10^{-4} \left( \frac{\rho}{2 {\rm \ g}/{\rm cm}^3 } \right)^{\frac{1}{2}} \frac{{\rm rad}}{\rm s}
.
\label{scdef}
\end{equation}

\noindent{}In the Solar system, the 2.33 hour critical limit corresponds to a density of about 2 g/cm$^3$.

\subsection{Main sequence evolution}  \label{sub3}

The Solar system is roughly 4.5 Gyr old, and contains a 
$\sim 5 \times 10^{-4} M_{\oplus}$ asteroid belt that has been 
continually ground down from a $\sim M_{\oplus}$ primordial belt 
\citep{botetal2005} through dynamical processes such as collisional 
evolution \citep{obrsyk2011}.  The most asymmetric asteroids
with orbits close to the Sun will have already been destroyed by rotational overspinning.  In fact, the 
YORP spin timescale is shorter than the spin timescale induced by collisions 
for asteroids in the Main Belt with sizes $\lesssim 10$ km
(\citealt*{maretal2011} and equations 2-3 of \citealt*{jacetal2014}).
Some asteroids have changed their shape due to collisions, eventually becoming asymmetric and hence destroyed.
Despite these possibilities, some asteroids will likely survive for the remaining main sequence lifetime of the Sun.
Extrasolar systems may exhibit the same characteristics, perhaps typically featuring asteroid belts 2-3 orders of
magnitude greater in mass than the Solar System's \citep{debetal2012}.  Our focus here is on the population of
asteroids which survive until the beginning of their star's giant branch phases.

\begin{figure*}
\centerline{
\psfig{figure=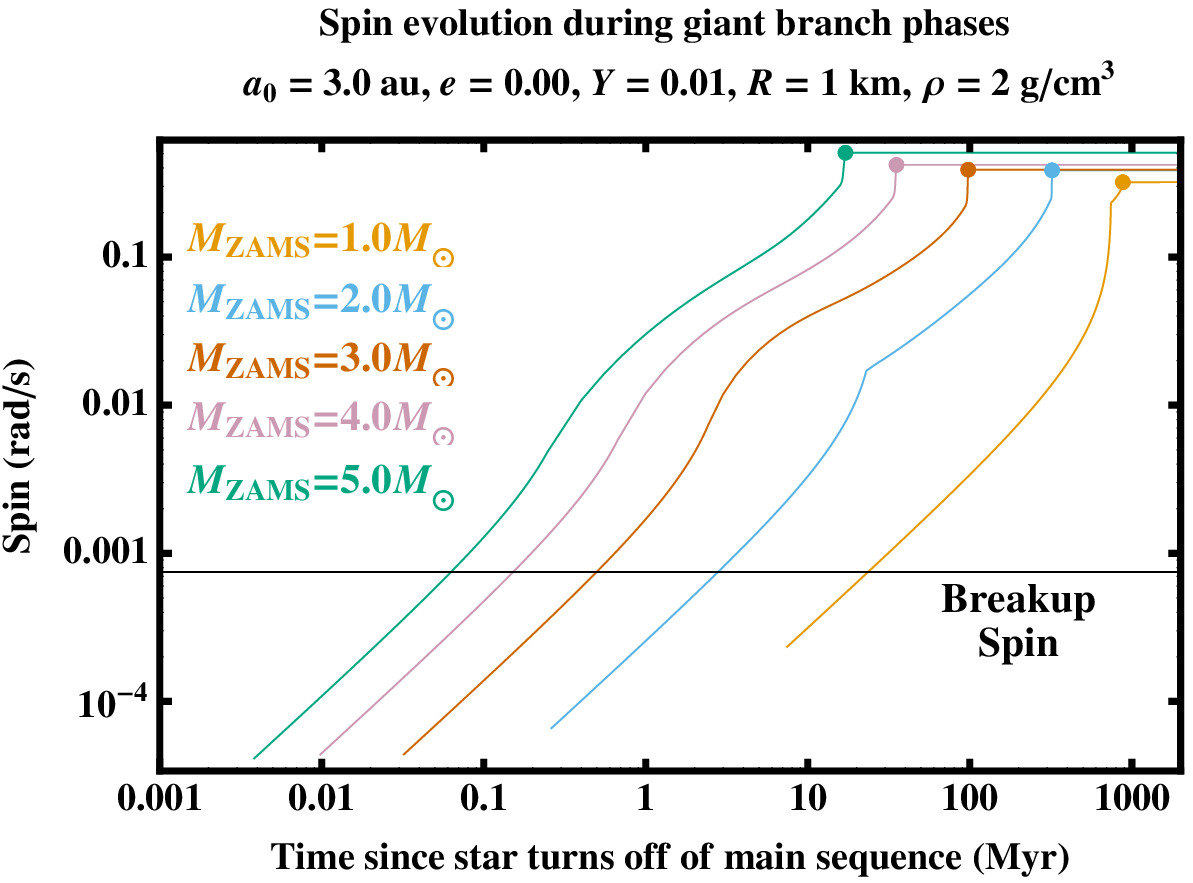,height=6.0cm,width=8.2cm} 
\psfig{figure=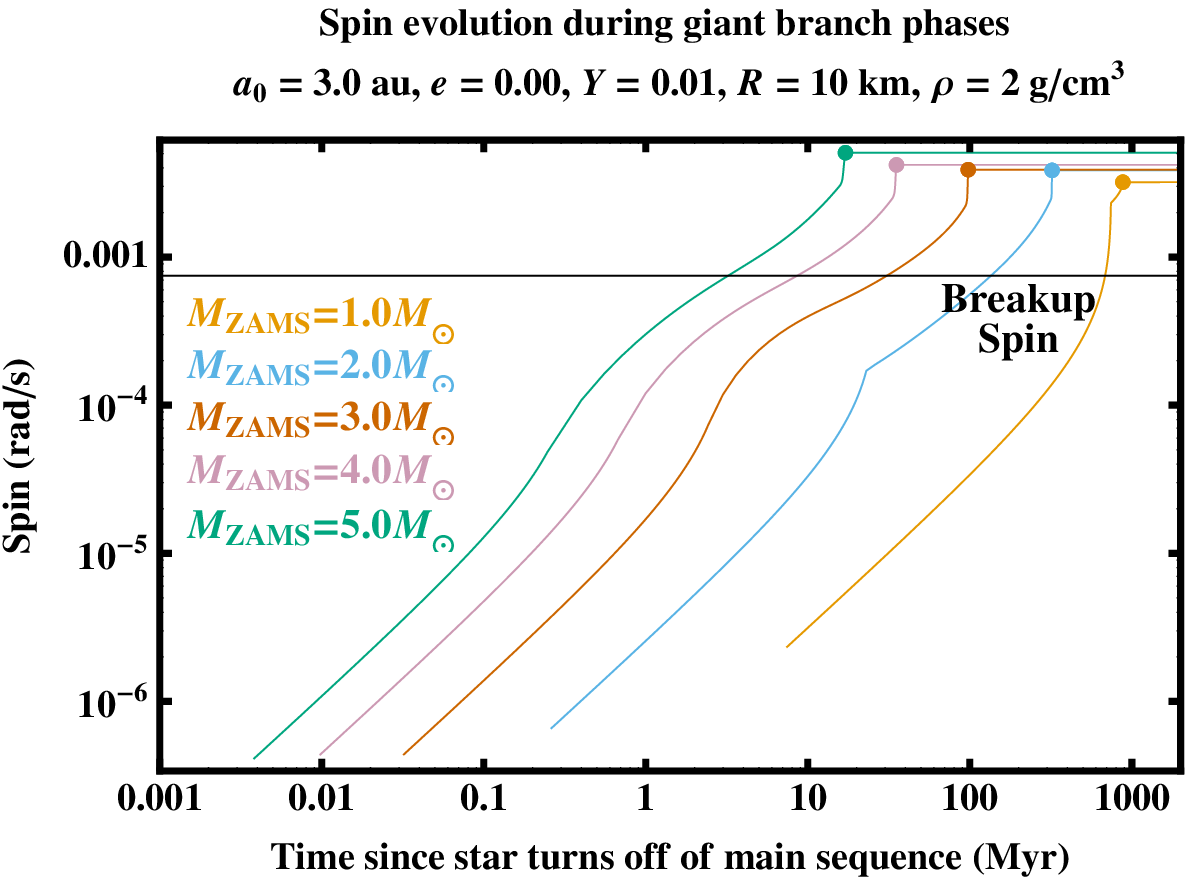,height=6.0cm,width=8.2cm}
}
\centerline{}
\centerline{
\psfig{figure=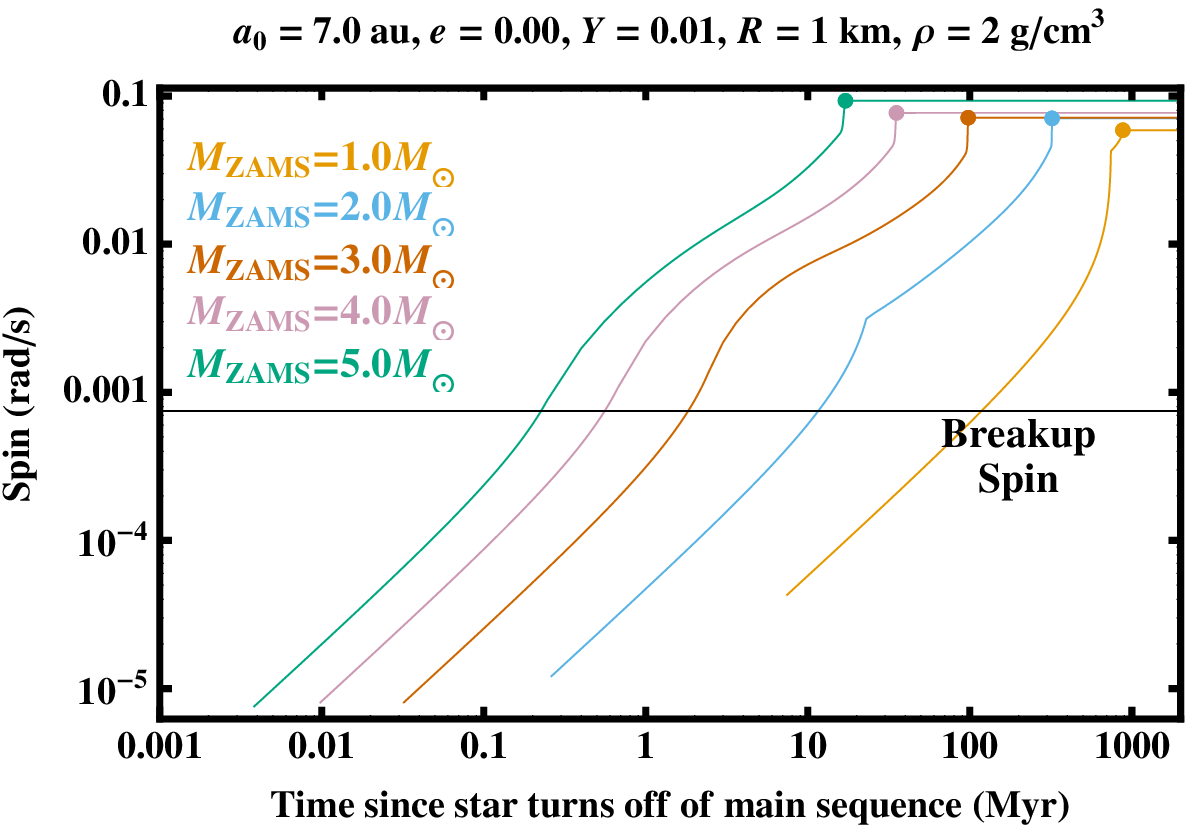,height=6.0cm,width=8.2cm} 
\psfig{figure=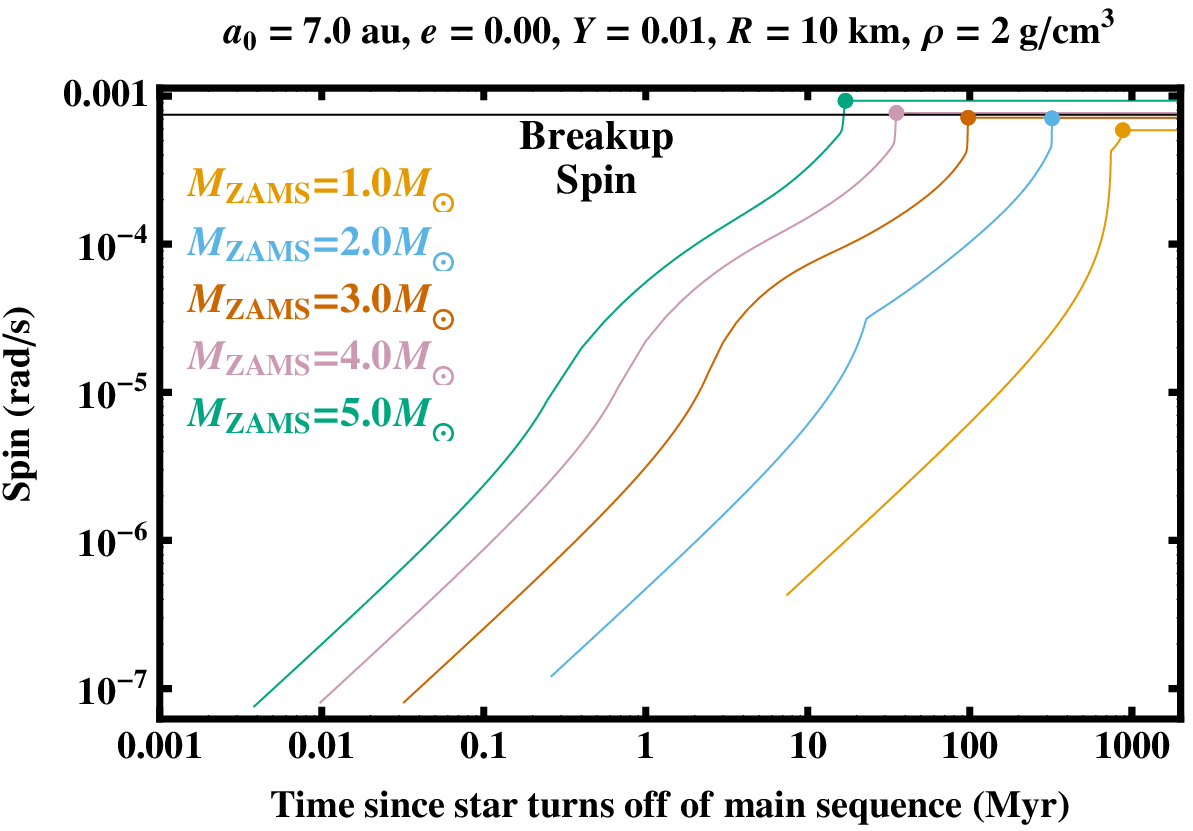,height=6.0cm,width=8.2cm}
}
\caption{Exo-asteroid belts will be destroyed during giant branch evolution.  The left and right
panels respectively feature the spin-up of objects with radii of $R=1,10$ km, and the top and bottom panels
respectively place these bodies on {\it initial} $a_0 = 3.0$, $7.0$ au circular orbits ($e=0$).  The objects 
have typical asteroid densities of $\rho = 2$ g/cm$^3$ and a degree of asymmetry corresponding to $Y=0.01$. The horizontal black lines
represent the critical spin value at which an asteroid will tear itself apart (equation \ref{scdef}).  
All objects are assumed to begin life on the giant branch with no spin, and the curves are drawn
from the beginning of the red giant branch phases. Dots indicate when the stars become WDs. The stars are all assumed to 
initially harbour Solar metallicity.}
\label{fisfig1}
\end{figure*}

\subsection{Giant branch stellar evolution} \label{sub4}

As the star evolves off the main sequence, asteroid orbits that are within a few hundred au will evolve adiabatically 
\citep{veretal2011}, meaning that the semimajor axis will expand but the
eccentricity will remain constant.  Mass emanating from the star during the giant branch phases will impact
orbiting asteroids \citep[see equations 2-3 of][]{dunlis1998} and perhaps change their shape.  However,
without a detailed model of impact, the character of this change is unknown, and hence we fix $Y$ as a constant
\footnote{In reality, variations in asteroid shape could occur on shorter timescales than the YORP timescale,
and the interplay between the two can significantly influence whether fission occurs \citep{cotetal2013}. 
Additionally, exo-meteoroid impacts on asteroids could enhance or mitigate the YORP effect \citep{wiegert2014}.}.
Mass clumps accompanying superwinds on the asymptotic giant branch may destroy particularly small asteroids 
outright.  In this vein, the effect of anisotropic
stellar mass loss may be more important for destruction of small bodies than for their orbital variations
from adiabaticity \citep{veretal2013a}.  The smallest bodies, with radii below about 1 cm, will be entrained
by the stellar wind \citep{donetal2010}.

In order to estimate the timescale for asteroidal break-up due to overspinning, we must consider 
the form of $L_{\star}(t)$.  
The evolution of the luminosity of a star is a complex function of its physical properties from the zero-age
main sequence (ZAMS) epoch, and hence are not amenable to simple functional forms.  Therefore, we compute 
$L_{\star}(t)$ along the giant branch phases by using the $\tt SSE$ stellar
evolution code \citep{huretal2000} with a standard Reimers mass loss coefficient of 0.5
on the red giant branch and the semi-empirical mass-loss rate of \cite{vaswoo1993} along the asymptotic giant branch phases.
All simulations assume Solar metallicity at the ZAMS.

\subsubsection{Stellar tides}

Small bodies which lie too close to an evolving star will be engulfed by the expanding
envelope.  The envelope expands to a distance in au roughly equal to the initial number of
Solar masses contained within the star \citep[Fig. 2 of][]{veretal2013b}.  However,
stellar tides extend beyond the reach of the star, as will be true with likely fatal
consequences for the Earth \citep{schcon2008}.  \cite{musvil2012} quantified this
extended reach along the giant branch phases by modelling how the stellar surface pulses.
Their Figure 7 plots the orbital distance of the initially most distant planet engulfed
by the bloated star as a function of stellar mass for $1M_{\odot} - 5M_{\odot}$.
We use their values for terrestrial planets as a proxy to guide our choices for our
initial pericentre $[\equiv a(1-e)]$ such that our asteroids will survive engulfment
along the giant branch phases.

\subsubsection{Wind drag}

Even if a small body survives engulfment and withstands impact from the strong stellar
wind, its outward motion due to stellar mass loss might be counterbalanced by drag
forces within the wind itself \citep{bonwya2010,donetal2010}.  These authors show that inward drift due to 
drag requires knowledge of the wind speed, is strongly dependent on the body size, and 
predominantly occurs when mass loss is strong.  However, they do not consider how 
asymmetry affects the inward motion.  Nevertheless, any type of inward drift of 
asymmetric asteroids will shorten the rotational disruption timescale.  We do not
model wind drag, and in effect then place a conservative upper limit on this timescale
by considering only outward motion due to mass loss.

\subsubsection{The Yarkovsky effect}

Different from YORP, the Yarkovsky effect is the orbital recoil experienced by a body
-- even if symmetric -- due to thermal radiation, and acts primarily on bodies
with sizes between 1 m - 100 m for the relevant time periods and distances in our Solar 
System \citep[e.g.][]{faretal1998}.  Like YORP, the Yarkovsky effect
has been empirically verified \citep[e.g.][]{rubincam1987}, and is a detailed function
of parameters such as the temperature, albedo, shape and size of the body.  The resulting drift
in the body can be towards or away from the star.  Solar system asteroids with radii under
100 m experience a typical maximum drift of about $10^{-3}$ au per Myr \citep{faretal1998}.

Here we consider only larger objects, and hence avoid the detailed modelling which would be 
necessary to understand the influence of the Yarkovsky effect.  Nevertheless, we point
out that the orbital evolution of these smaller asteroids may be dramatically affected
by re-radiation.  For Solar system asteroids, the insolation due to Yarkovsky is proportional to the 
orbital acceleration and hence the time evolution of orbital parameters such as the semimajor axis 
\citep[Section 3.1 and equation 42 of][]{botetal2000} and eccentricity.
Consequently, for extrasolar systems, the contribution to the asteroid's motion includes
a multiplier $k = L_{\star}/L_{\odot}$ in the equations for $da/dt$ and $de/dt$, just as 
for equation (\ref{sdef}).  By using {\tt SSE}, we find that for main sequence stellar masses of 
$\left\lbrace 1M_{\odot},2M_{\odot},3M_{\odot},4M_{\odot},5M_{\odot},6M_{\odot},7M_{\odot},8M_{\odot}\right\rbrace$, the maximum factor increase in luminosity ($k_{\rm max}$) is
$\approx \left\lbrace 4070, 9200, 18700, 31800, 46300, 61800, 70000, 92000\right\rbrace$.
Consequently, the maximum possible drift due to Yarkovsky is about $10^{-6}$ au per yr
for the lowest mass stars and $10^{-4}$ au per yr for the highest mass stars.  These values
are comparable to the orbital expansion rates due to mass loss at the tip of the asymptotic
giant branch.

\subsection{White dwarf stellar evolution} \label{sub5}

When the star becomes a WD, its luminosity profile changes, quickly becoming sub-Solar.
The time since becoming a WD is known as the {\it cooling age}.  The cooling ages
at which WD luminosities become sub-Solar is a few Myr in all cases.  After another 10 Myr,
the luminosity decreases by an order of magnitude.  For cooling ages of 100 Myr, $k \sim 0.01$.
Consequently, any asymmetric asteroids which have survived to this stage are likely to remain intact
for several Gyr of WD evolution unless they are already spinning near the critical rate.

\subsection{Simulations}  \label{sub6}

Here we perform simulations that showcase the extent of rotational disruption
when the parent star leaves the main sequence.
We assume, conservatively, that all asteroids begin evolving during giant branch
evolution without any spin and that the asteroid shape ($Y = 0.01$) does not change
The eccentricity of the asteroid remains constant, as adiabatic mass loss predicts.  
Both the luminosity of the star and the semimajor axis of the asteroid are varied; the latter is varied adiabatically and so is independent of initial orbital orientation.
We model the evolution of stars with main sequence progenitor masses of $1-5M_{\odot}$,
as these likely represent the range of progenitor masses which have yielded the vast majority of currently observed
WDs.

We present results in Fig. \ref{fisfig1}, which highlights the great extent
of radiation-induced destruction for any exo-asteroid belts.  Even slightly asymmetric
asteroids as large as 10 km at 7 au on the main sequence will likely reach the disruption
limit.  The higher the stellar mass, the faster the disruption occurs.  Asteroids
which fail to disrupt before the WD is formed are likely to remain intact orbiting
the WD (in the absence of other influences), as extending the plots to include
WD luminosity cause the curves to nearly flatline.

\begin{figure}
\centerline{
\psfig{figure=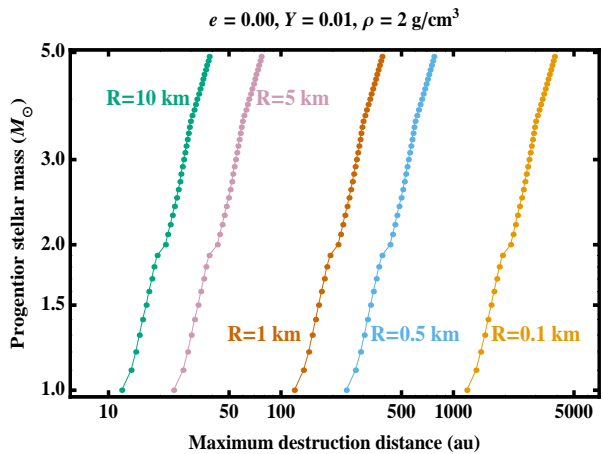,height=6.0cm,width=8.3cm} 
}
\caption{The maximum (final) distance, or semimajor axis, at which asteroids with forever circular orbits ($e=0$), a radius $R$, density of $\rho = 2$ g/cm$^3$ and asymmetry parameter $Y$ of $0.01$ can be destroyed by YORP.  Each dot represents the result of an integration initialised with the highest initial semimajor axis for which destruction occurs. Because mass loss becomes nonadiabatic at a few hundred au (Veras et al. 2011), and this motion is not modelled here, the two rightmost curves are more approximate than the other three. 
}
\label{maxreach}
\end{figure}

Consequently, we predict that WD systems can be split into three regions
with regard to populations of objects with radii of about 100m-10km:
(1) an inner region, which is devoid of such asteroids, (2) an intermediate region, which
is strewn with debris from rotational breakup, and (3) an outer region, which contains
predominantly fully-intact asteroids.  The boundaries between the regions are strongly
dependent on the ZAMS stellar mass; the boundary between the second and third regions 
are shown in Figure \ref{maxreach}, which plots the maximum possible distance at which 
asteroids can be broken apart by YORP.  The figure illustrates that this destruction boundary 
ranges from a few tens of au to a few thousands of au depending on asteroid radius.
Broken-up asteroids could then provide a widely-extended debris
reservoir of pollutants with a particular size distribution \citep[e.g.][]{wyaetal2014} 
to be delivered to the WD via dynamical interactions
with planets.  These results could also have implications for debris discs which are observed around subgiant and giant stars \citep{bonetal2013,bonetal2014}.

\section{Summary}

We have identified a potentially significant source of debris in post-main-sequence planetary
systems: the remains of asymmetric asteroids which have spun up beyond their breaking points due to
stellar radiation during the giant branch stages of stellar evolution.  Typical asymmetric asteroids 
with radii between about 100m - 10km that reside within about 7 au of the star during the main sequence will
be destroyed, leaving a debris field orbiting WDs at distances which can range from a few tens to a few thousands of au.  
Objects larger than 10km in size will be largely unaffected by YORP.  The majority of the asteroidal-based mass 
would be contained in these large objects, such as analogues of Ceres and Vesta.
The debris from rotational breakup may provide a reservoir for WD disc creation and ultimately atmospheric pollution.  WD luminosity
itself is too low to destroy asteroids in timescales within Gyr unless the asteroids are already spinning near the breakup speed.

\section*{Acknowledgments}

The research leading to these results has received funding from the European 
Research Council under the European Union's Seventh Framework Programme (FP/2007-2013) 
/ ERC Grant Agreement n. 320964 (WDTracer).  

\label{lastpage}
\end{document}